\documentclass[aps,prb,showpacs, groupedaddress,twocolumn, superscriptaddress]{revtex4}
\usepackage{amsmath,graphicx,latexsym,times,color}
\usepackage{setspace}
\usepackage{hyperref}
\usepackage{array}
\usepackage{titlesec}

\begin{document}

\title{Symmetry-selected spin-split hybrid states in C$_{60}$/ferromagnetic interfaces}

\author{Dongzhe Li}
\affiliation{Service de Physique de l'Etat Condens\'e (CNRS UMR 3680), IRAMIS/SPEC, CEA Saclay, Universit\'e Paris-Saclay, F-91191 Gif-sur-Yvette Cedex, France}

\author{Cyrille Barreteau}
\affiliation{Service de Physique de l'Etat Condens\'e (CNRS UMR 3680), IRAMIS/SPEC, CEA Saclay, Universit\'e Paris-Saclay, F-91191 Gif-sur-Yvette Cedex, France}
\affiliation{DTU NANOTECH, Technical University of Denmark, {\O}rsteds Plads 344, DK-2800 Kgs. Lyngby, Denmark}

\author{Seiji Leo Kawahara}
\affiliation{Laboratoire Mat\'eriaux et Ph\'enom\`enes Quantiques, Universit\'e Paris Diderot et CNRS, UMR 7162, Case courrier 7021, 75205 Paris Cedex 13, France}

\author{J\'er\^ome Lagoute}
\affiliation{Laboratoire Mat\'eriaux et Ph\'enom\`enes Quantiques, Universit\'e Paris Diderot et CNRS, UMR 7162, Case courrier 7021, 75205 Paris Cedex 13, France}

\author{Cyril Chacon}
\affiliation{Laboratoire Mat\'eriaux et Ph\'enom\`enes Quantiques, Universit\'e Paris Diderot et CNRS, UMR 7162, Case courrier 7021, 75205 Paris Cedex 13, France}

\author{Yann Girard}
\affiliation{Laboratoire Mat\'eriaux et Ph\'enom\`enes Quantiques, Universit\'e Paris Diderot et CNRS, UMR 7162, Case courrier 7021, 75205 Paris Cedex 13, France}

\author{Sylvie Rousset}
\affiliation{Laboratoire Mat\'eriaux et Ph\'enom\`enes Quantiques, Universit\'e Paris Diderot et CNRS, UMR 7162, Case courrier 7021, 75205 Paris Cedex 13, France}

\author{Vincent Repain}
\affiliation{Laboratoire Mat\'eriaux et Ph\'enom\`enes Quantiques, Universit\'e Paris Diderot et CNRS, UMR 7162, Case courrier 7021, 75205 Paris Cedex 13, France}

\author{Alexander Smogunov}
\email{alexander.smogunov@cea.fr}
\affiliation{Service de Physique de l'Etat Condens\'e (CNRS UMR 3680), IRAMIS/SPEC, CEA Saclay, Universit\'e Paris-Saclay, F-91191 Gif-sur-Yvette Cedex, France}

\date{\today}

\begin{abstract}
The understanding of orbital hybridization and spin-polarization at the organic-ferromagnetic interface is essential  in the search for efficient hybrid spintronic devices. Here, using first-principles calculations, we report a systematic study of spin-split hybrid states of C$_{60}$ deposited on various ferromagnetic surfaces: bcc-Cr(001), bcc-Fe(001), bcc-Co(001), fcc-Co(001) and hcp-Co(0001). 
We show that the adsorption geometry of the molecule with respect to the surface crystallographic orientation of the magnetic substrate as well as the strength of the interaction play a crucial role in the spin-polarization of the hybrid orbitals. We find that a large spin-polarization in vacuum above the buckyball can only be achieved if the molecule is adsorbed  upon a bcc-(001)  surface by its pentagonal ring. Therefore, bcc-Cr(001), bcc-Fe(001) and bcc-Co(001) are the optimal candidates. Spin-polarized scanning tunneling spectroscopy measurements on single C$_{60}$ adsorbed on Cr(001) and Co/Pt(111) also confirm that both the symmetry of the substrate and of the molecular conformation have a strong influence on the induced spin polarization. Our finding may give valuable insights for further engineering of spin filtering devices through single molecular orbitals.
\end{abstract}


\newcommand{\Ea}{\ensuremath{E_a}}
\newcommand{\Eb}{\ensuremath{E_b}}
\newcommand{\Ec}{\ensuremath{E_c}}
\newcommand{\Ed}{\ensuremath{E_d}}
\newcommand{\Ee}{\ensuremath{E_e}}

\newcommand{\eV}{\ensuremath{\,eV}}

\maketitle

\section{ Introduction }
\label{Intro}

The organic spintronics has become an exciting research field in nanoelectronics because of its flexibility, low production costs and easy functionalization. In particular, the carbon-based materials are promising candidates for efficient tunnel barrier in spintronics devices (an organic layer sandwiched between two magnetic electrodes) with a large spin-dependent transport due to its low intrinsic spin-orbit coupling as well as weak hyperfine interaction. Unexpected large magnetoresistances and a large spin-dependent transport length have been reported in spin valves using Alq$_3$ \cite{Xiong-2004, Sun-2010, Barraud-2010}, carbon nanotubes \cite{Tsukagoshi-1999}, self-assembled molecular wires \cite{Petta-2004} and C$_{60}$ \cite{Zhang-2013} as an organic spacer layer between two magnetic layers. Recently,  spin coherent transport in spin valves based on C$_{60}$ at room temperature \cite{Gobbi-2011} was also reported. 

In this context, the study of organic-ferromagnetic hybrid interfaces at the molecular level plays a key role to get a better understanding of the physical mechanism involved in spin injection and the subsequent spin-transport in organic spin valves\cite{Sanvito-2010}. Due to the numerous peculiar properties of C$_{60}$ (high symmetry, easy production, thermal and mechanical resilience etc), a large effort has been dedicated to the survey of  C$_{60}$-ferromagnetic interfaces. For instance, the strong hybridization between the $d$ states of magnetic surfaces and the lowest unoccupied molecular orbitals (LUMOs) of a C$_{60}$ molecule leads to spin-polarized molecular states close to the Fermi level in C$_{60}$/Cr(001) \cite{Alex-2012}, C$_{60}$/Ni(111) \cite{Yoshida-2013} and C$_{60}$/Fe(001) \cite{Deniz-2014-C60} interfaces. 

The present work is motivated by our recent study of single C$_{60}$ molecules adsorbed on a Cr(001) surface at the pentagon site (lowest energy configuration) \cite{Alex-2012}. There, large tunneling magnetoresistances (TMR) were observed by Spin-Polarized Scanning Tunneling Spectroscopy (SP-STS). Performing  {\it ab initio} calculations we have attributed the TMR effect to the spin-splitting of one of the three LUMO orbitals close to the Fermi energy due to strong hybridization with the magnetic substrate. This orbital,  labeled $m=0$, is  localized at the pentagon center and extends significantly in the vacuum. On another side, the two other LUMO orbitals, labeled $m=\pm 1$, are strongly localized on the pentagon ring and are almost not detectable in experiment.

One can question, therefore, about the generality of such results and under which conditions (substrate nature, crystallographic orientation, C$_{60}$ adsorption site etc.) they can be reproduced or even optimized. In this paper we will thus address two main aspects: i) the strength and details of the spin-polarization of LUMO orbitals due to magnetic substrate and ii) the detectability of such spin-polarized orbitals (if any) in SP-STS measurements which would give rise to the TMR effect. 

Our manuscript is organised as follows. In Sec. \ref {method}, we present the computational and experimental methods used in this work. In Sec. \ref {Results and discussion}, we present a systematic $\textit {ab initio}$ study and show SP-STS measurements on C$_{60}$ adsorbed on two different substrates. 
A particular emphasis will be made on the spin-resolved local density of states (LDOS) in the vaccum above the molecule
which is the relevant quantity when comparing to STM spectra. Finally, the conclusions will be drawn in Sec. \ref{conclusions}.

\label{table_energy}
\begin{center}
\begin{table*}[htbp]
\scalebox{1.0}{
 \begin{tabular}{cccccccccc}
&  & & \includegraphics[scale=0.06]{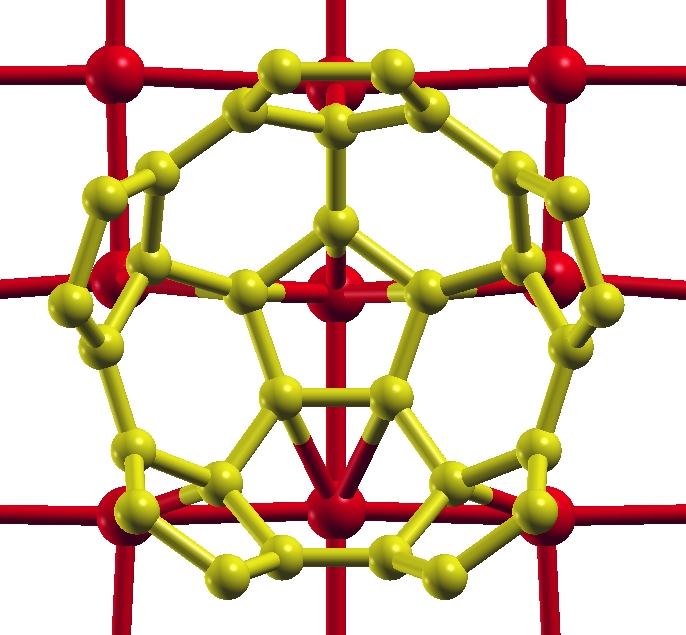} &\includegraphics[scale=0.06]{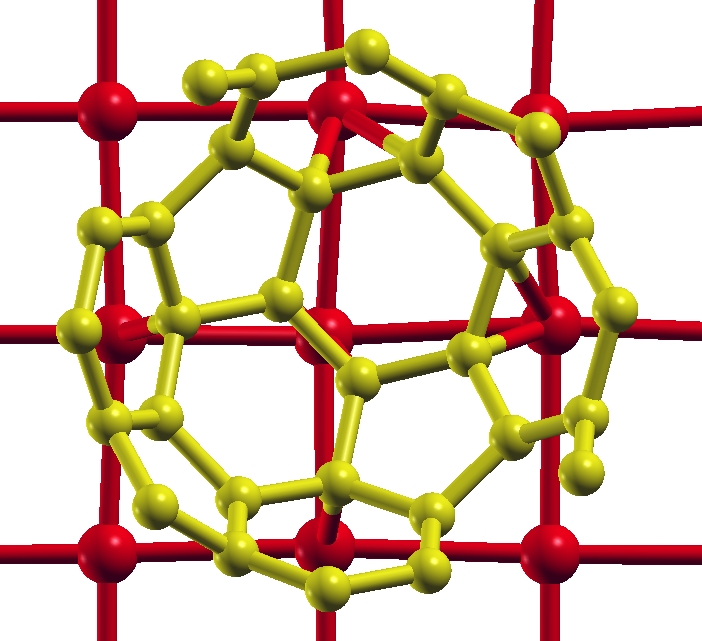} &  & & \includegraphics[scale=0.062]{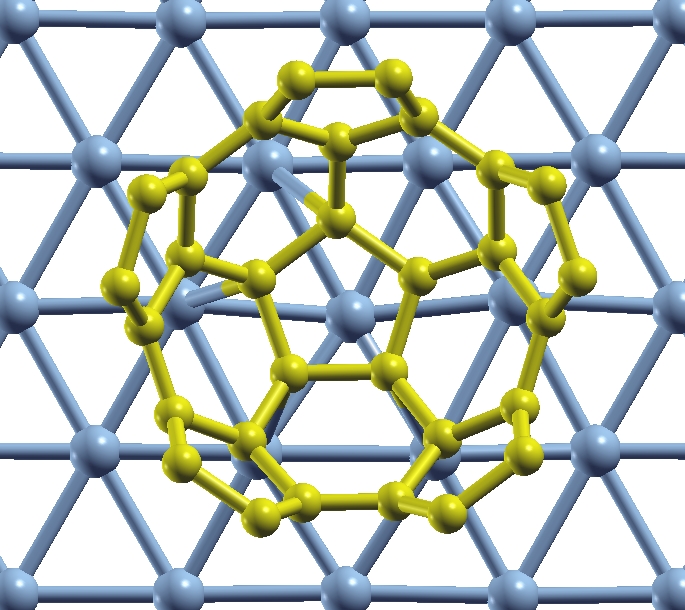} &\includegraphics[scale=0.061]{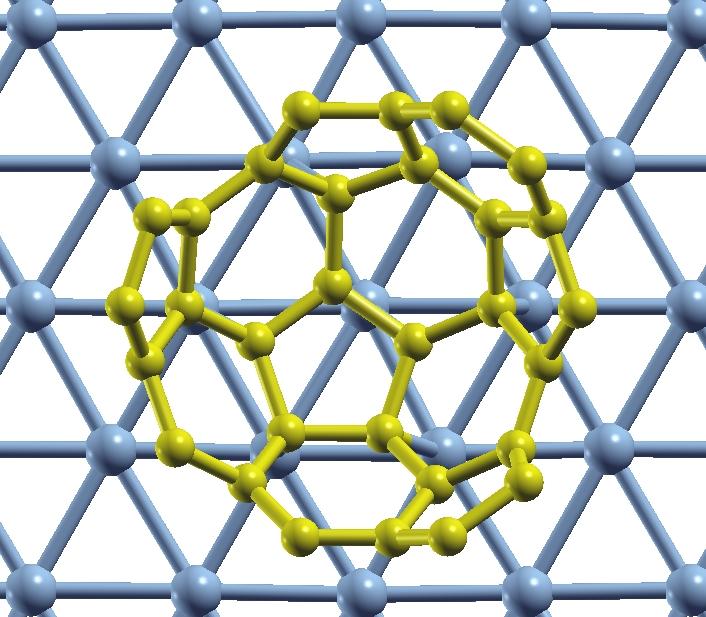} &  \includegraphics[scale=0.065]{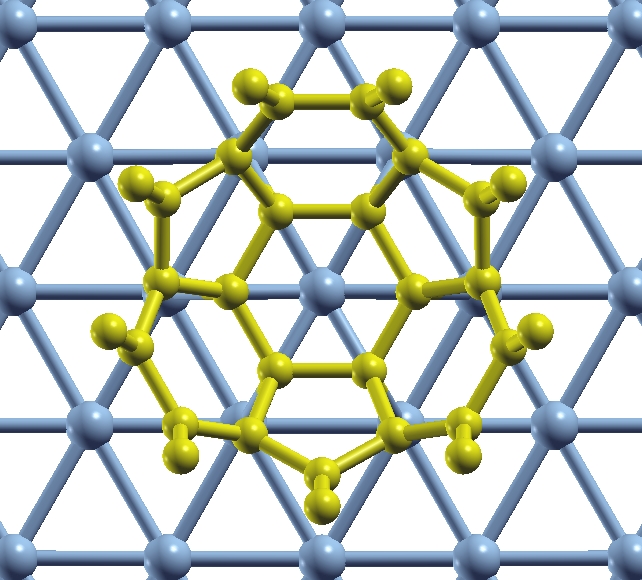}\\
\hline\hline
\multicolumn{1}{c|}{} & \multicolumn{2}{c|}{bcc-Cr(001)} & \multicolumn{2}{c|}{bcc-Fe(001)} & \multicolumn{1}{c|}{bcc-Co(001)} & \multicolumn{1}{c|}{fcc-Co(001)}& \multicolumn{3}{c}{hcp-Co(0001)}\\
\multicolumn{1}{c|}{$$} & \multicolumn{1}{c} {Pentagon}\cite{Alex-2012} & \multicolumn{1}{c|} {6:6 bond} & \multicolumn{1}{c} {Pentagon} & \multicolumn{1}{c|} {6:6 bond} & \multicolumn{1}{c|} {Pentagon}& \multicolumn{1}{c|} {Pentagon} & \multicolumn{1}{c} {Pentagon} &  \multicolumn{1}{c}  {5:6 bond} & \multicolumn{1}{c} {Hexagon} \\ \hline
$E_\textnormal{b}$ (eV) & $3.90$ & $3.50$ & $2.32$ & $2.71$  &$2.90$&$2.29$ & $0.79$ & $1.35$ & $1.21$\\
$d_{\textnormal{C-Substrate}}$  (\AA)  & $2.05$ & $2.08$  & $2.05$  & $2.01$  &$1.98$&$1.95$& $2.08$  & $1.95$ & $2.08$\\
Charge transfer (e) &$0.50$& $0.36$ &$0.38$& $0.32$ &$0.30$&$0.03$ &$0.00$& $0.02$ & $0.08$\\
$M_s$ on C$_{60}$ ($\mu_\text{B}$) &$-0.45$& $-0.03$ &$-0.60$& $-0.38$  &$-0.45$ &$-0.40$&$-0.31$& $-0.23$ & $-0.33$\\
\hline\hline
\end{tabular}}
\caption {Binding energy $E_\text{b}$,  shortest carbon-metal interatomic distance, charge transfer to the molecule and induced spin moment $M_s$ on C$_{60}$ adsorbed on cubic magnetic surfaces [bcc-Cr(001) \cite{Alex-2012}, bcc-Fe(001), bcc-Co(001) and fcc-Co(001)], and on a hexagonal magnetic surface [hcp-Co(0001)] with different adsorption geometries. Inequivalent adsorption geometries considered in the paper are demonstrated on upper panels where only the lower half of
C$_{60}$ and the surface layer of the substrate are shown for simplicity.}
\label{table_energy}
\end{table*}
\end{center}

\section{ Methodology}
\label{method}

\subsection{ Density functional theory calculations}
\label{DFT}

Spin-polarized $\textit {ab initio}$ studies were carried out using the plane wave electronic structure package Quantum ESPRESSO \cite{Giannozzi2009} in the framework of the density functional theory (DFT). Generalized gradient approximation in Perdew, Burke and Ernzerhof parametrization \cite{pbe-1996} was used for the exchange-correlation functional within the ultra-soft pseudopotential formalism. Energy cut-offs  of 30 Ry and 300 Ry were employed for the wavefunctions and the charge density, respectively. 
The interfaces were modeled by five layer slabs of magnetic material with a $(4 \times 4)$ in-plane periodicity  on which  one C$_{60}$ molecule was deposited.  The super-cell periodicity was increased up to $(5 \times 5)$  in the case of fcc-Co(001) and hcp-Co(0001) to avoid interaction between the buckyballs. 
In the ionic relaxation, the Brillouin-zone has been discretized by using a $(4 \times 4 \times 1)$ $k$-points mesh and a Marzari Vanderbilt cold smearing parameter of 0.01 Ry.  The two bottom layers were fixed while other three layers of substrate and C$_{60}$ molecule were relaxed until the atomic forces are less than 0.001 eV/\AA. In addition, a vacuum space of about 20 \AA~was taken to separate two neighboring slabs in the $z$ direction (perpendicular to the surface) in order to avoid unphysical interactions. The electronic structure of the relaxed structures has been studied by using a denser $(6 \times 6 \times 1)$ $k$-points mesh and two additional atomic layers were added.

\subsection{ Experiment}
\label{STM}

The experiments have been performed in an ultra-high vacuum setup (P$<$10$^{-11}$~mbar), with a Omicron low-temperature STM working at 4.6~K. To obtain a Co close-packed surface, we have deposited by e-beam evaporation a submonolayer Co coverage on a Pt(111) surface at room temperature. The Pt(111) substrate has been cleaned by Ar ions sputtering cycles (1~kV, P = 5.10$^{-6}$~mbar) at 800~K followed by flash annealing to 880~K until the Auger spectrum does not show any carbon contaminant. This system is known to display spin contrast among the different Co islands, blocked either in a spin up or spin down out-of-plane configuration\cite{meier_spin-dependent_2006}. The spin polarized results have been obtained using a freshly annealed tungsten tip coated with around 20~nm of iron. Once the spin contrast has been achieved, the C$_{60}$ has been deposited at 4.6~K. As deposited, the adsorption configurations of the C$_{60}$ are numerous, even though a vertex configuration seems to be more frequent (cf. appendix C). Such a sample makes difficult an analysis of the spin contrast for different molecules on the same geometry. We have therefore annealed the sample to room temperature, what leads to single C$_{60}$ molecules adsorbed on a pentagonal configuration (cf appendix C).The Cr(001) sample has been cleaned by Ar ions sputtering cycles (2~kV, P = 5.10$^{-6}$~mbar) at 850~K. The C$_{60}$ has been deposited at 4.6~K onto the sample and directly measured, what allows to observe various adsorption geometries. It is worth noting that an annealing at room temperature of such a sample leads to a favored pentagonal configuration that has already been studied in detail and display a spin-split LUMO level\cite{Alex-2012}.

\begin{figure}[!htbp]
\centering
\includegraphics[scale=0.4]{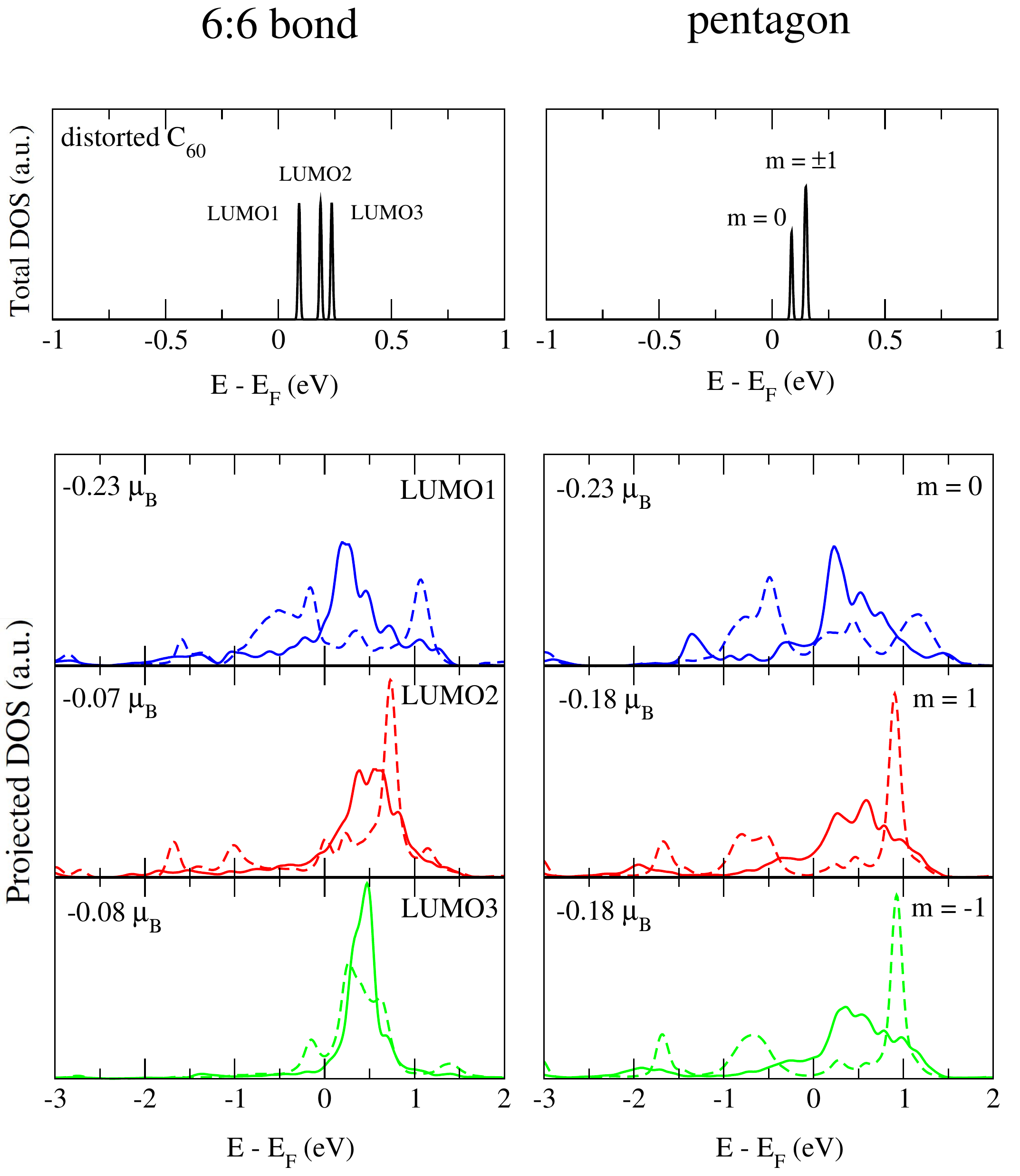}
\caption{\label{LDOS_Fe001}
(Color online) 
Electronic structure of a C$_{60}$ molecule absorbed on bcc-Fe(001) by a pentagon face and in 6:6 bond geometry: 
(a) upper panels shown the total DOS of the free (but distorted) C$_{60}$ molecule. Lower panels show the spin-resolved density of states of the full C$_{60}$/Fe system projected on the three LUMO levels of the isolated molecule;  
(b)  present the spin-resolved vacuum LDOS at 5 \AA~above the free (distorted) molecule (upper panels) and above the  C$_{60}$ in contact with the bcc-Fe(001) substrate (lower panels). Spin up/down curves are plotted by solid/dashed lines.}
\end{figure}

\begin{figure}[!htbp]
\centering
\includegraphics[scale=0.5]{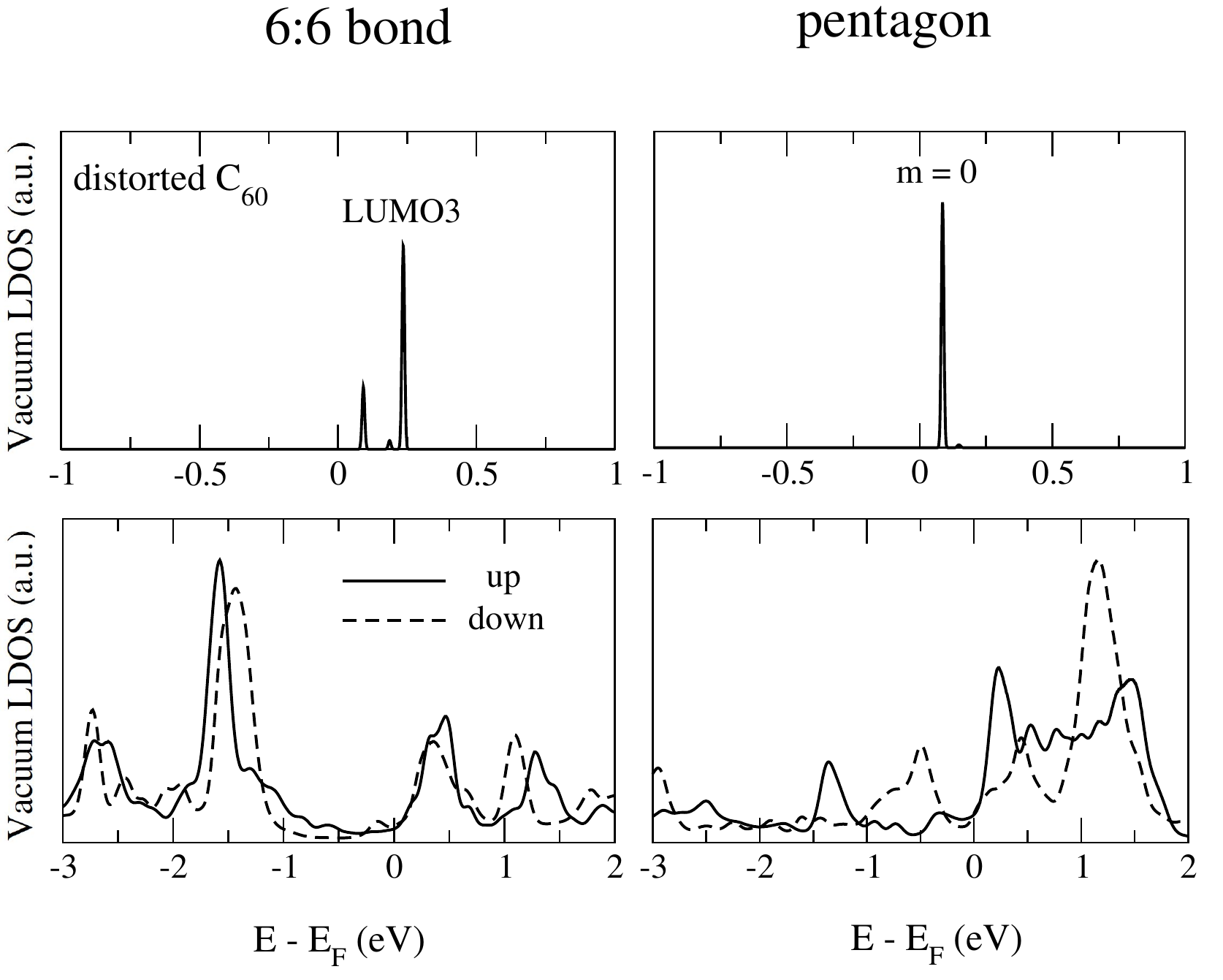}
\caption{\label{LDOS_Fe001_2}
(Color online) 
Electronic structure of a C$_{60}$ molecule absorbed on bcc-Fe(001) by a pentagon face and in 6:6 bond geometry: 
(a) upper panels shown the total DOS of the free (but distorted) C$_{60}$ molecule. Lower panels show the spin-resolved density of states of the full C$_{60}$/Fe system projected on the three LUMO levels of the isolated molecule;  
(b)  present the spin-resolved vacuum LDOS at 5 \AA~above the free (distorted) molecule (upper panels) and above the  C$_{60}$ in contact with the bcc-Fe(001) substrate (lower panels). Spin up/down curves are plotted by solid/dashed lines.}
\end{figure}

\section{Results and discussion}
\label{Results and discussion}

\subsection{DFT calculations }

We have investigated various C$_{60}$/ferromagnetic interfaces, i.e., different chemical elements (Cr, Fe, Co) and crystalline structures (cubic and hexagonal) of the substrates as well as different molecule adsorption sites (pentagon, hexagon, 5:6 bond, 6:6 bond), with a particular emphasis on the hybridization between molecular levels and surface spin-polarized states. In Table \ref{table_energy} the main results are summarized for all studied configurations. As a general result, we mention the antiferromagnetic (with respect to the substrate) 
magnetization induced on C$_{60}$ molecules and, on another side, the decrease of the spin moment for the surface atoms beneath the molecule due to hybridization with molecular 
states (see Appendix A for more details) -- these effects have been already reported for Cr\cite{Alex-2012} and Fe\cite{Deniz-2014-C60}.    

\subsubsection{C$_{60}$ on  Fe bcc-(001) surface geometry}

We start our discussion with the case of a C$_{60}$ molecule adsorbed on a Fe-bcc(001) surface which we expect to behave similarly to the Cr-bcc(001) since both metals have the same crystalline structure and almost the same lattice parameter. We have considered two adsorption geometries presented in Fig.~\ref{LDOS_Fe001}.
In contrast to the case of  Cr, for which the pentagonal configuration is the most stable, the lowest energy configuration for C$_{60}$/ Fe(001)  corresponds to the molecule bound by a 6:6 bond to the iron surface which agrees with previous DFT calculations.\cite{Tran-2013} 
The pentagon geometry has a higher energy of about $0.4$ eV (see Table \ref{table_energy}). 

In Fig. \ref{LDOS_Fe001} (a), for both adsorption geometries, we present in upper panels the density of 
states (DOS) for the isolated (but distorted, as in the contact with substrate) molecule. Lower panels show the DOS 
for the full C$_{60}$/Fe system projected onto LUMO orbitals of the (distorted) C$_{60}$ molecule which largely dominate around the Fermi energy. Note that due to the structural distortion, caused by the interaction with the substrate, there is a breaking of the icosahedral symmmetry and the original three-fold degenerate C$_{60}$ LUMOs  will split differently for the two adsorption geometries.

In the ideal pentagonal geometry the LUMO states can be labeled by an integer $m$ that reflects their symmetry with respect to the five-fold rotation axis passing through the centers of two opposite pentagonal facets.\cite{Manousakis-1991} For deposited molecule this symmetry axis is lost  and the only remaining symmetry is the reflection plane perpendicular to the surface and dividing the molecule into two. As a consequence, the three-fold degenerate LUMO levels labeled by $m=0$ and $m=\pm 1$ are split into one ``even'' $m=0$ and two (almost) degenerate $m=\pm 1$  (one ``odd'' and one ``even'') levels. 
The influence of the substrate can evidently not be be summed up by a simple lifting of the disctete levels of the molecule: they will also be strongly broadened (and spin-split) due to their hybridization with the spin-polarized substrate states. Nevertheless, it is interesting to formally divide the coupling process in two contributions: a first one related to the distorsion of the molecule and another one to the electronic hybridization.
The splitting of the LUMO levels described above is clearly seen, the $m=0$ being lower in energy. But the hybridization produces much more drastic effect on the levels of the molecule which are broadened over an energy range of more than 3 eV. The two types of LUMOs are affected rather differently but still bear some resemblances: all of them are negatively polarized, the  spin down states show a resonance below the Fermi level (around -0.5 eV) while spin up states have a resonance slightly above the Fermi level. These features  seem to be more pronounced, however, for the $m=0$ states.

For the 6:6 bridge geometry, symmetry arguments put forward for the pentagonal geometry do not apply anymore since no symmetry is present at all. The three LUMO states, (labeled LUMO1, LUMO2, and LUMO3), are fully split as seen from the DOS of the isolated deformed molecule. However the LUMO1 seems to keep a dominant $m=0$ character while $m=\pm 1$ mix strongly. As a result the PDOS of LUMO1 resembles somewhat the one obtained with the pentagonal geometry while LUMO2 and LUMO3 are rather different and show almost no exchange splitting. 

The PDOS analysis, presented above, is very instructive but since one of our goals is to predict which system will show spin-split states measurable in a tunneling regime, typically probed by a spin-polarized scanning tunneling spectroscopy (SP-STS), the so-called vacuum local density of states (LDOS)  is a more appropriate quantity. Indeed, in a simplified Tersoff-Hamann approach \cite{PRL-Wortmann},  the spin-polarized differential conductance is  simply related to the spin-resolved LDOS of the sample at the STM tip position:
\begin{equation}
G=\frac{dI}{dV}\propto  \sum_{\sigma}n_{\rm T}^{\sigma}n_{\rm S}^{\sigma}(R_{\rm T}, E_F+eV)
\end{equation}
where $n_{\rm T}^{\sigma}$, $n_{\rm S}^{\sigma}(R_{\rm T}, E_F+eV)$ are, respectively, the spin-dependent tip DOS (assumed to be constant in energy) and vacuum LDOS of the sample (C$_{60}$ molecule deposited on surface) calculated at the tip position $R_{\rm T}$ above the molecule and at the energy corresponding to applied voltage $V$. 

We present in Fig. \ref{LDOS_Fe001} (b) the vacuum LDOS calculated by integrating 
$n_{\rm S}^{\sigma}(R_{\rm T}, E)$ inside a small cubic box of size 0.4 \AA~at 5~\AA~above the C$_{60}$ molecule for both spin polarizations. For isolated molecules (upper panels), it allows to single out molecular orbitals 
with the slowest decay in vacuum which should be accessible to the STM measurements.
It turns out that, for both adsorption configurations, only one of the LUMO orbitals will mainly dominate in the tunneling current.
In the pentagon geometry, it is $m=0$ orbital which is essentially located in the 
center of the pentagon as will be discussed later.
In the 6:6 bond configuration, it is LUMO3 orbital while the detection of the
LUMO1 orbital (resembling the $m=0$) is strongly unfavored    
due to the rotation of the molecule.
As a consequence, the vacuum LDOS for adsorbed molecules (lower panels) follows very closely the corresponding molecular PDOS around the Fermi level. Since, in the pentagon geometry, the $m=0$ is clearly spin-split this results in large spin-polarization of vacuum LDOS leading to large magnetoresistance which should be visible in SP-STS experiment, similar to what has been measured on the Cr surface\cite{Alex-2012}.
In contrast, for the 6:6 bridge geometry, the LUMO3 orbital shows almost no exchange splitting so the vacuum LDOS is rather spin-independent which should translate into very small TMR values.  
Note also that in this bridge configuration a specific feature in LDOS is seen at -1.5 eV. 
It can be attributed to the HOMO levels which in addition show a rather modest spin-splitting. These states are hardly visible in the case of the pentagonal geometry since the electronic density of the HOMO states is rather low at the center of the pentagon. 

One can thus conclude that even though both adsorption configurations 
possess spin-polarized LUMO-derived states around the Fermi energy it is not enough to get a spin-polarized signal in a SP-STS measurement. 
It is demonstrated by the bridge geometry where the slowest decaying state  in vacuum  (and thus dominating the tunneling current) was found to be only weakly spin-polarized.     
Finally, we note that we have also looked at the bridge adsorption site on the Cr surface and have not found neither any significant spin-polarization of the vacuum 
LDOS (see Appendix B, Fig. \ref{bilan_mag_surfaces} (d)).  

\begin{figure*}[!htbp]
\centering
\includegraphics[scale=0.38]{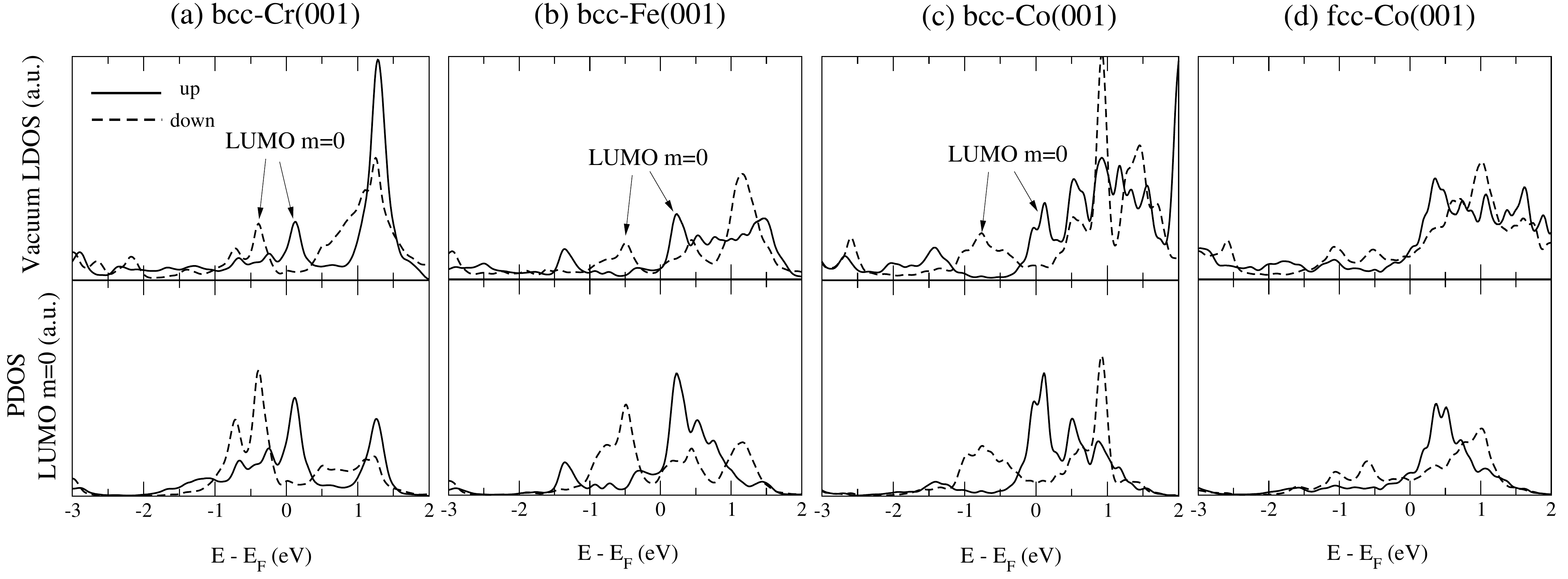}
\caption{\label{LDOS_pentagon}
Electronic structure of a C$_{60}$ molecule adsorbed  on top of cubic magnetic surfaces in pentagonal ring geometry: (a) bcc-Cr(001) \cite{Alex-2012}, (b) bcc-Fe(001), (c) bcc-Co(001) and (d) fcc-Co(001). Upper panels: Spin-resolved vacuum local density of states (LDOS) at 5 \AA~above the C$_{60}$ molecule. Lower panels: Density of states of the C$_{60}$/substrate projected onto the molecular orbital LUMO m = 0 of the deformed molecule. Spin up/down curves are plotted by solid/dashed lines.}
\end{figure*}

\subsubsection{C$_{60}$ on various $(001)$ surfaces in pentagonal geometry}

Since the pentagonal adsorption site on a cubic lattice seems to be the best configuration to generate a strong exchange splitting of the LUMO states that can be probed in the tunneling regime, it is tempting to look at other possible cubic substrates such as Co-bcc(001) and Co-fcc(001) and compare them with Fe and Cr cases discussed above.
For the sake of completeness we present in Fig.~\ref{LDOS_pentagon} the vacuum LDOS and PDOS on the $m=0$ LUMO -- which plays a crucial role --  for all the four cases together.
As previously mentioned,  the Cr and Fe bear some similarities. As shown in Fig. \ref{LDOS_pentagon}, in the vicinity of the Fermi level the vacuum LDOS is largely dominated by the LUMO states of $m=0$ character which are at the origin of two resonances: 
one at $-0.5$ eV for the down spin and another one slightly above the Fermi level (0.2 eV) for the up spin polarization. 
There is also another sharp feature above 1 eV but that do not show any clear spin-splitting and comes probably from LUMO+1 states having somewhat similar shape as the $m=0$ LUMO orbital. The case of bcc-Co(001) is also quite similar: two spin-split LUMO $m=0$ resonances are present approximately at the same energy position (the spin down feature being slightly shifted towards lower energy: -0.8 eV).  Above the Fermi energy the broad feature extending towards high energy originates from LUMO+1 derived states.
In contrast, the fcc-Co(001) shows a rather different behavior.  In particular, there is a low and featureless electronic density below the Fermi level (down to at least -3 eV below the Fermi level) and a modest exchange splitting above the Fermi level. 
The departure from the case of the bcc-(001) substrates can probably be attributed to the densest atomic packing  of the fcc(001)  surface: the interatomic distance of the 2D square lattice of fcc-Co(001) ($a=2.49$ {\AA}) is much smaller than the one of  bcc-Co(001) ($a=2.84$ {\AA}). Therefore, more carbon  atoms of the C$_{60}$ molecule are involved in a bonding with the substrate but the strength of the hybridization due to a given metal-carbon  bond is weaker. It results in an overall binding energy similar to the case of the bcc-Co(001) surface (see Table I) but since the vacuum LDOS is still largely dominated by the contribution of the $m=0$ molecular orbital (which is less hybridized) a smaller spin-splitting is found. 
We can summarize that the spin-splitting of the $m=0$ LUMO becomes less and less pronounced when passing from the bcc-Cr to fcc-Co accompanied by a continuous disappearance of related spin-split peaks in the vacuum LDOS.

\begin{figure}[!htbp]
\centering
\includegraphics[scale=0.36]{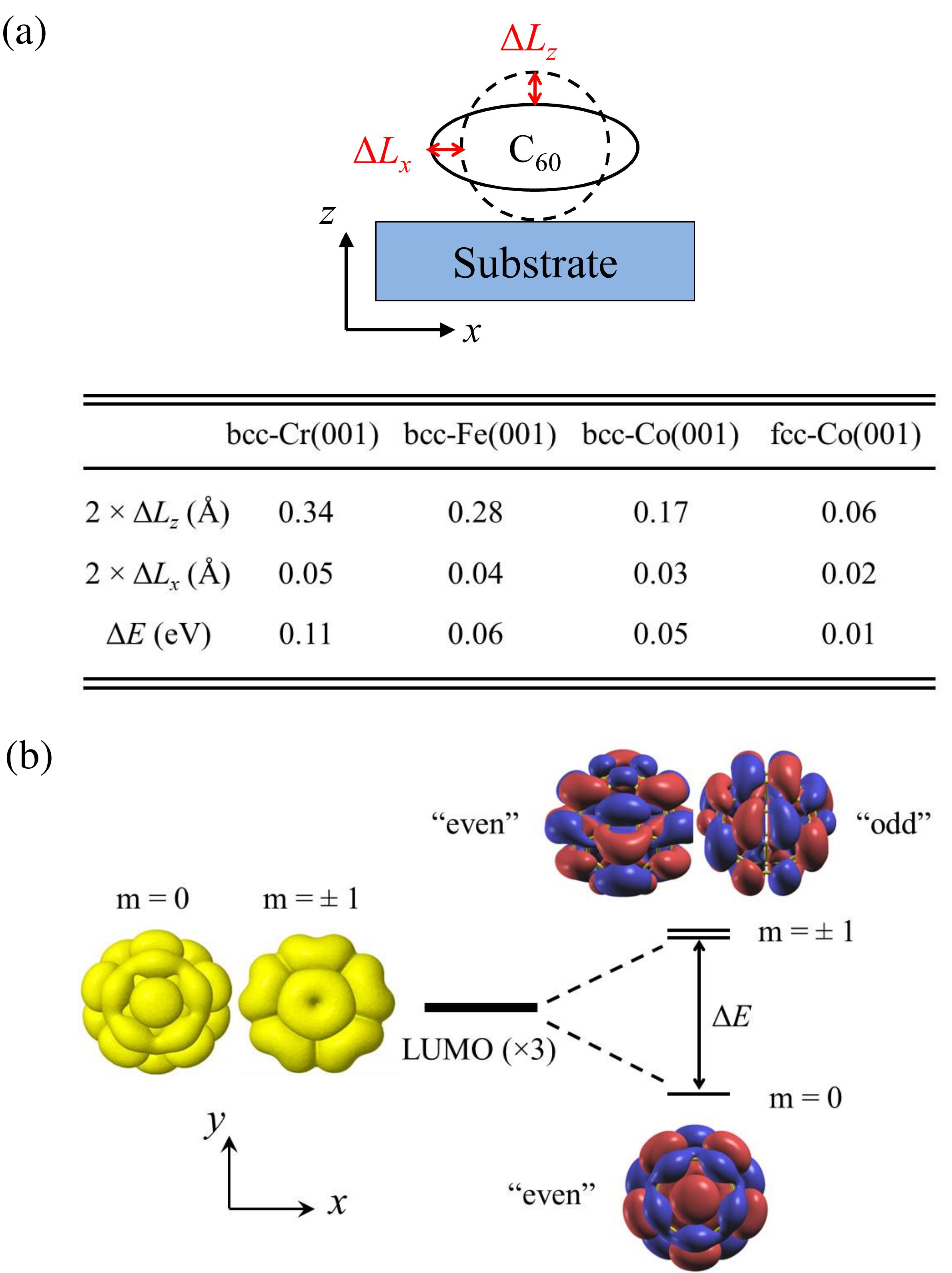}
\caption{\label{Distortion}
(Color online) (a) C$_{60}$ deformation due to the interaction with a $(001)$  surface of cubic materials.  $\Delta L_z$   ($\Delta L_x$) quantifies the deformation of the adsorbed molecule (with respect to the free molecule) in the direction $z$ perpendicular ($x$ parallel) to the surface. A schematic illustration of the deformation is also presented on the top panel.   (b) Energy level diagram of LUMO levels for the free but distorted C$_{60}$ molecule (after ionic relaxation of the full system). The 3-fold degeneracy of the C$_{60}$ LUMOs is lifted by the molecule deformation into a single $m=0$ state and two almost degenerate $m=\pm1$ states which can be classified as "even" or "odd" with respect to the mirror plane $yz$. The spatial representation of  the three LUMO levels  is also presented, the isosurfaces of positive and negative values are shown in red and blue.}
\end{figure}

To further rationalize this behavior we have tried to quantify the molecule distortion and relate it to the LUMO splitting in the four cases discussed above (Fig. \ref{Distortion}). The deformation of the molecule in contact with the substrate is evaluated by comparison with the free molecule in a direction perpendicular and  parallel to the surface, namely, $\Delta L_z$ and $\Delta L_x$, respectively. As shown in Fig. \ref{Distortion} (a), the flattening of the molecule ($\Delta L_z$) is much more important than its lateral  expansion ($\Delta L_x$) which can be neglected. Interestingly, the deformation is basically  proportional to the binding energy (see Table \ref{table_energy}) for the bcc surfaces while almost no distortion is found for the fcc-Co surface although the interaction is substantial. Once again this is probably due to the more dense packing of the surface atomic layer of the fcc structure. 

The level splitting of three-fold degenerate LUMO levels  goes in line with the amplitude of the deformation as demonstrated by Fig. \ref{Distortion} (b). The main out-of-plane deformation $\Delta L_z$, preserving the five-fold rotational axis, will lift the degeneracy between $m=0$ and $m=\pm 1$ orbitals, with the gap $\Delta E$
roughly proportional to $\Delta L_z$.  The $m=0$  has a circular shape extending out of the molecule at the center of the 
pentagon while $m=\pm 1$ states have a strong localization on the pentagon ring.
The additional small (and slightly anisotropic, $\Delta L_x \approx \Delta L_y$)  in-plane distortion, and the electronic coupling to the cubic substrate will break the five-fold rotational symmetry leaving only mirror symmetry with respect  to the plane $yz$ (see Fig. \ref{Distortion} (b)) passing through the molecule and perpendicular to the surface.
The $m=0$  is  of ``even'' symmetry while $m=\pm 1$ states will be slightly split into one state of ``odd'' and one of ``even'' symmetry. The important point is that this latter ``even'' state can admix to the $m=0$ since they have the same symmetry, destroying its circular shape, and the amount of this admixture depends on the gap $\Delta E$ -- the bigger $\Delta E$ the smaller the admixture.

Since the $m=0$ state is crucial in producing the spin-split features in vacuum LDOS, as has been shown above, it is crucial to keep it as pure as possible. Therefore, the larger $\Delta L_z$ the more pronounced are the two spin-split peaks in the vacuum LDOS which explains the general tendency observed  in Fig.~\ref{LDOS_pentagon}.

\begin{figure}[!htbp]
\centering
\includegraphics[scale=0.5]{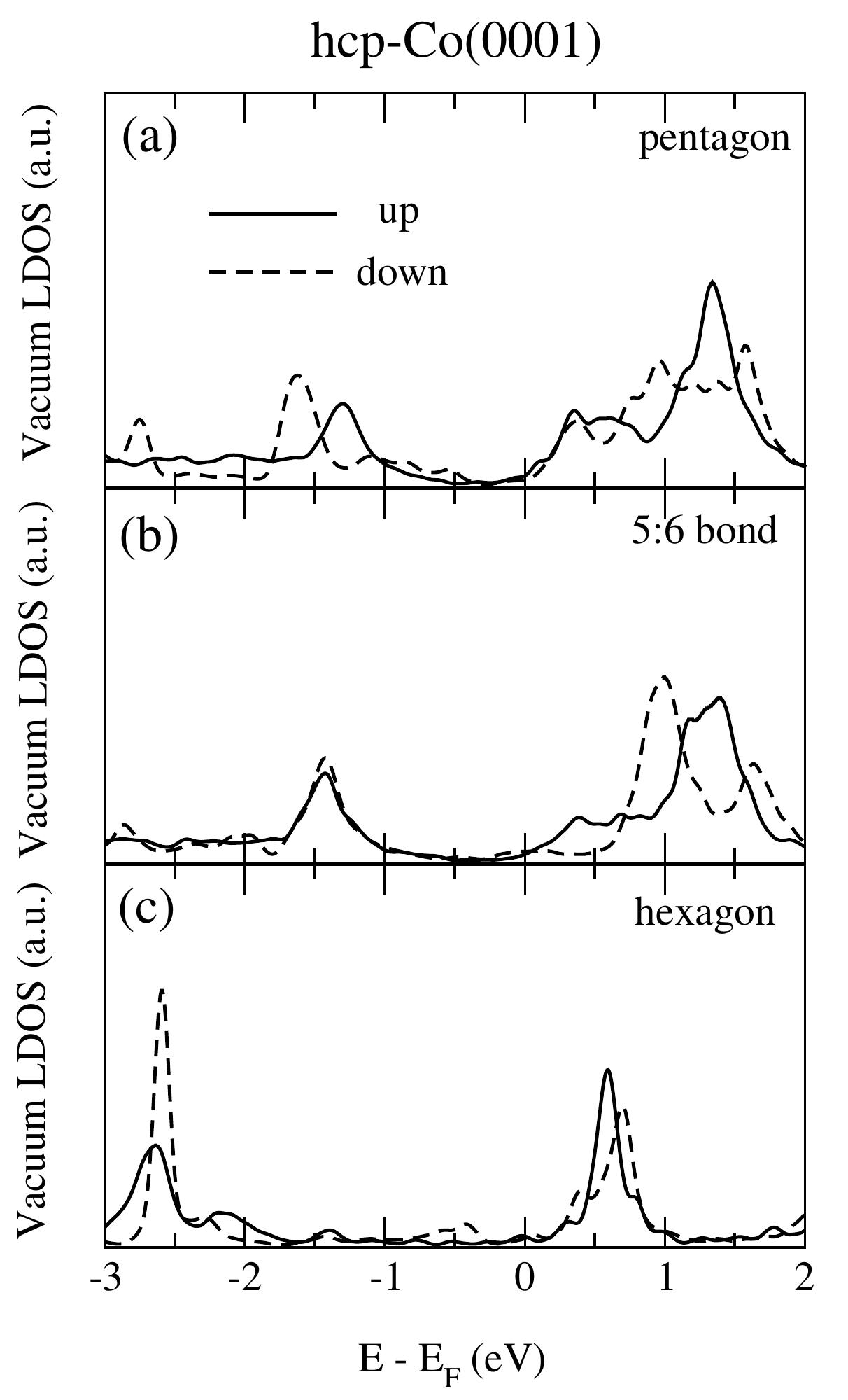}
\caption{\label{LDOS_Co_0001}
Calculated spin-resolved vacuum LDOS at 5 \AA~above the C$_{60}$ molecule deposited on hcp-Co(001) surface with various adsorption  sites: (a) a pentagon ring , (b) a 5:6 bond, (c) a hexagon ring. Spin up/down curves are shown by solid/dashed lines.}
\end{figure}

\subsubsection{ C$_{60}$ on hexagonal magnetic surface }
\label{Co(0001)}
Finally let us consider another important substrate with a different symmetry:  hcp-Co(0001). 
The most stable relaxed atomic configuration of the C$_{60}$ was found to be the 5:6 bond (see Table I) where the molecule is bound to a surface Co atom by a pentagon-hexagon edge. Nevertheless since it is rather easy for this system to be trapped in a local metastable minimum we have considered the two other configurations, namely pentagonal and hexagonal. 
In Fig. \ref{LDOS_Co_0001}, we plot the vacuum LDOS of C$_{60}$ absorbed on hcp-Co(0001) for these three different adsorption sites. 
As a general result, in all three cases the vacuum LDOS is very low around the Fermi level and show almost no spin-polarization.
Interestingly, in the pentagonal geometry the PDOS analysis (see Supplementary materials) reveals again slight spin-polarization of the 
LUMO1 orbital around $-0.5$ eV.  However, due to symmetry mismatch with the substrate, all the LUMOs are mixed and decay almost equally to the vacuum which results in very weak spin-polarization of LDOS above the C$_{60}$ molecule. 
In the hexagonal geometry, on the contrary, only LUMO3 orbital has a large extension in the vacuum, but it is not spin-polarized
(see Supplementary materials) which results again in weak spin-polarization of the vacuum LDOS.  
We can conclude, therefore, that for observing a low-bias TMR signal in C$_{60}$/ferromagnet systems it is crucial to have only one 
low-decaying LUMO state which, in addition, should be spin-polarized due to hybridization with ferromagnetic substrate. 
This was found to happen only for $m=0$ LUMO orbital in the pentagon adsorption geometry on cubic magnetic substrates. 

It should be finally emphasized that if the contact regime is concerned (rather than the tunneling regime), which could be realized in spin valve devices using a ``sandwich" trilayer geometry, then the vacuum decay rate of different LUMO
orbitals is no longer of importance and all the LUMOs are expected to contribute to the current. In this case the standard spin-polarized PDOS analysis presented 
in Fig.~\ref{LDOS_Fe001} (a) for C$_{60}$/Fe system is already sufficient. In particular, it
suggests that in both C$_{60}$/Fe adsorption geometries a spinpolarized LUMO orbital shows up, so that the spin-polarized
current should be observed irrespective of the details in C$_{60}$/Fe interface—the result recently reported for the
Fe-bcc(001)/C60/Fe-bcc(001) junctions \cite{Deniz-2014-C60}.

\subsection{SP-STS measurements }

\begin{figure}[!htbp]
\centering
\includegraphics[scale=0.48]{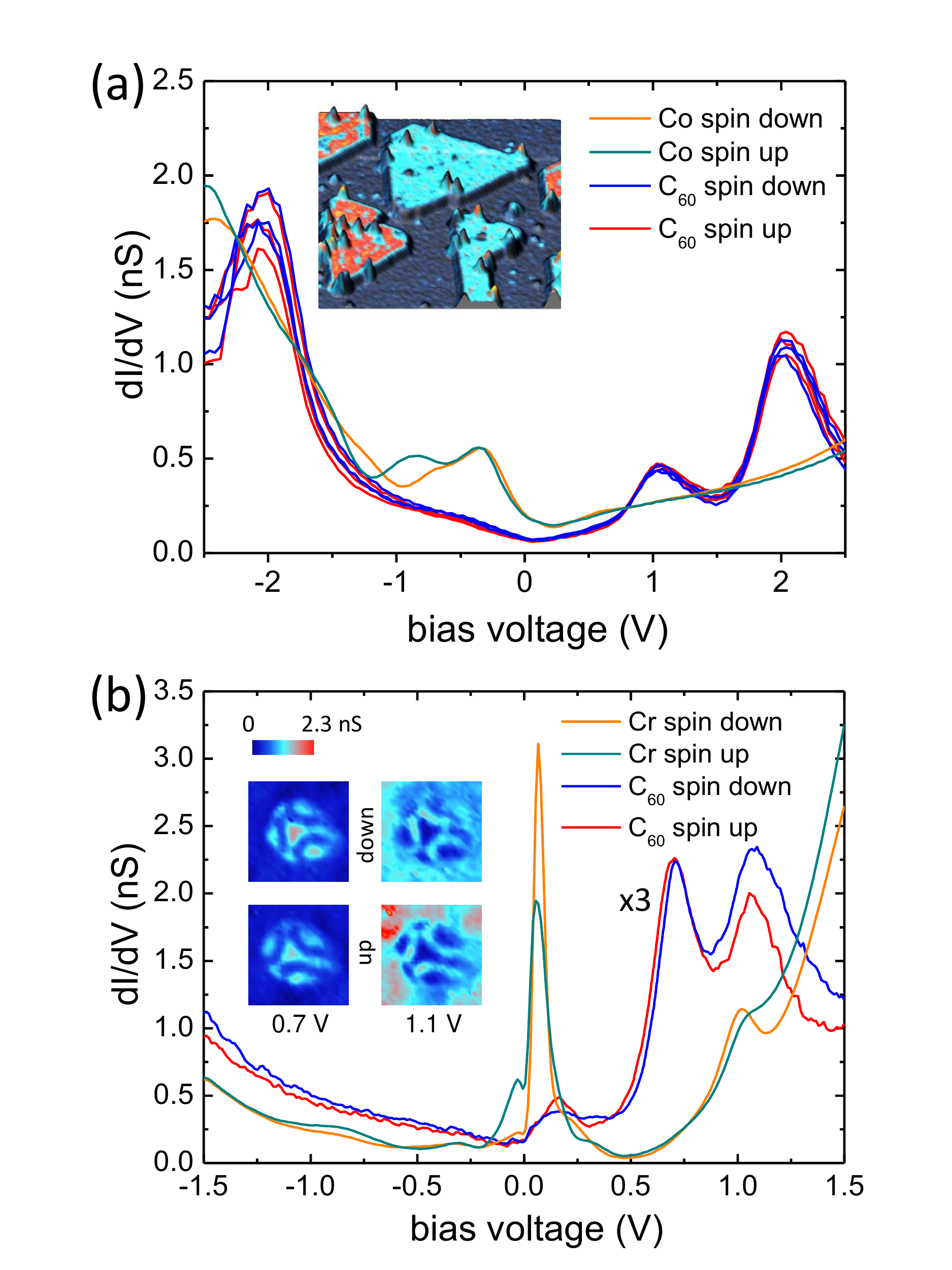}
\caption{\label{STS}
(Color online)  (a) Differential conductance as a function of the tunneling voltage  recorded on Co/Pt(111) and on C$_{60}$ adsorbed on a pentagonal configuration. Green and orange spectra are the reference taken on up spin and down spin Co islands. Blue and red spectra are taken on top of six different C$_{60}$ molecules adsorbed respectively on up spin and down spin Co islands. In inset, a 3D topographic STM image (50 $\times$ 50~nm$^2$) of C$_{60}$ molecules adsorbed on Co islands on Pt(111). The color code is the conductance at -1~V, showing the spin contrast in the Co islands. (b) Differential conductance as function of the tunneling voltage recorded on Cr(001) and on C$_{60}$ adsorbed on an hexagonal configuration. Orange and green spectra display the signal of two adjacent Cr terraces, i.e. down spin and up spin. The blue and red curves show spectra averaged over a whole C$_{60}$ molecule adsorbed on down spin and up spin terrace respectively. For clarity, those latter curves have been magnified by a factor three. In inset, conductance images of the corresponding C$_{60}$ molecules, taken at the energies of the molecular levels (0.7 and 1.1~V).}
\end{figure}

In order to confirm that the spin polarization probed by SP-STM above a C$_{60}$ molecule is very sensitive to the substrate crystal structure and to the adsorption geometry of the molecule, we have measured single C$_{60}$  adsorbed on Co/Pt(111) and on Cr(001). It is worth noting that it is generally difficult to achieve a quantitative comparison between experimental and theoretical results for several reasons. Firstly, the experimental systems can display complex structures (discommensuration lines on Co/Pt(111), cf. appendix C) and local defects that cannot be taken into account in calculations as they would require very large unit cells and a too long computational time. Secondly, the LDOS calculation in vacuum on a single point (Tersoff-Hamann model) does not give always an accurate description of STM results. Molecular orbitals are generally broader in experiments than calculated and at different energies\cite{lu_charge_2004}. It should be even worth for SP-STM where the tip polarization is likely to play an important role in the quantitative analysis of spin polarized spectra. Fig.~\ref{STS} (a) shows the results for C$_{60}$ on Co/Pt(111). The image in inset shows several one monolayer high Co islands with a color code displaying their differential conductance at -1~V which is known to display a spin contrast\cite{meier_spin-dependent_2006}. The spectra averaged over two opposite spin direction islands (yellow and green) show no feature above $E_F$ and broad peaks in the negative range, from 0 to -1.3~V, that have been ascribed to electronic surface resonances of Co\cite{meier_spin-dependent_2006}. The blue and red spectra have been taken above six single C$_{60}$ molecules, all adsorbed in a pentagonal configuration, as checked by high resolution images, but in contact either with a spin down Co island (blue spectra) or with a spin up Co island (red spectra). All these six spectra are very similar and typical of C$_{60}$ adsorbed on a close-packed surface\cite{lu_charge_2004} with a peak associated to the HOMO level at -2~V, to the LUMO at 1.1~V and to the LUMO+1 at 2~V. The important result is that we cannot observe any significant difference between the molecules adsorbed on spin down and spin up islands. The comparison with theoretical calculations can only be qualitative as the real structure of the Co islands is very complex with different stacking area, discommensuration lines and surface dislocations\cite{meier_spin-dependent_2006}. However, the absence of measurable spin polarization is in good agreement with the conclusion of Sec.~\ref{Co(0001)} where all the adsorption geometries on Co(0001) show a very weak spin polarization.

Fig.~\ref{STS} (b) focus on the case of C$_{60}$ molecules adsorbed on a roughly hexagonal configuration on a Cr(001) surface. A closer look at high resolution images, such as images in inset, shows that the studied molecules are in between an hexagonal and a 6:6 adsorption geometry. The typical spectra of the raw Cr surface for spin down (yellow) and spin up terraces (green) show a sharp features close to the Fermi level that has been associated to a $s-p_z-d_{z^2}$ surface state and an electronic state around 1~V, both displaying spin contrast on the conductance images\cite{Lagoute-2011, Habibi-2013}. Two spectra associated with a molecule lying on a spin down terrace (up images in inset) and a molecule on a spin up terrace (down images in inset) are shown respectively in blue and red. These spectra are very different from the ones measured on molecules adsorbed in a pentagonal configuration\cite{Alex-2012}. More specifically, they do not show a spin-split LUMO orbital, in good agreement with the theoretical findings. Mainly two LUMO states at 0.7~V and 1.1~V are observed, with only a slight change in intensity for spin up and spin down configurations for the LUMO+1. This can be compared with the results of Fig.~\ref{bilan_mag_surfaces} (d) (Appendix B), having in mind that the experimental adsorption geometry is not exactly the same as the calculated one. The quantitative agreement is pretty good with a LUMO level insensitive to the spin polarization of the substrate and a LUMO+1 level that is slightly different for the spin up and spin down configurations.

Finally, those experimental measurements on Co(0001) and Cr(001) confirm that the vacuum LDOS above the C$_{60}$ molecules only shows a clear spin polarization and spin-split orbital in a pentagonal adsorption geometry on the (001) surface of a bcc. Other adsorption geometries or different substrates like Co(0001) are less favorable to observe such spin-split molecular orbitals.

\section{ Conclusions}
\label{conclusions}
To conclude, we have investigated systematically from first-principles, the spin-polarized hybrid states of C$_{60}$ deposited on ferromagnetic surfaces such as bcc-Cr(001), bcc-Fe(001), bcc-Co(001), fcc-Co(001) and hcp-Co(0001). As a general feature, a strong chemisorption of the buckyball on ferromagnetic surfaces leads to a remarkable drop of the spin moment for magnetic surface atoms and an induced negative spin moment for the C$_{60}$ molecule. It was found that the degree of spin-polarization of the C$_{60}$ LUMO depends strongly on both the adsorption geometry  and the symmetry of the surface. Due to symmetry matching between the C$_{60}$ LUMO states with the underlying surface and a non-negligible distortion in perpendicular direction to the surface, the 3-fold degeneracy of LUMOs of free C$_{60}$ could be lifted energetically and spatially. As a result, a large spin-polarization of a single hybrid orbital is only achieved if a C$_{60}$ molecule is adsorbed by a pentagon face on cubic surfaces, such as bcc-Cr(001), bcc-Fe(001) and bcc-Co(001), and is related to the spin-split $m=0$ LUMO which is strongly localized at the centre of the pentagon ring. 
In contrast, the adsorption on the hexagonal hcp-Co(0001) surface leads to very small vacuum molecular induced spin polarization. Our theoretical results are qualitatively confirmed by SP-STS measurements that show no measurable spin polarization for C$_{60}$ on Co/Pt(111) and no spin-split LUMO orbital for a non pentagonal adsorption geometry on Cr(001). Understanding the mechanism of spin-polarized hybrid states at the interface is an essential ingredient for the engineering of spin filtering in carbon-based spintronics devices and we expect that this work will help for their future rationalized design.

\section{Acknowledgement}
This work has been funded partly by ANR-BLANC-12 BS10 006 and by the HEFOR project of the Labex SEAM. This work was performed using HPC computation resources from GENCI-[TGCC] (Grant No. 2015097416).
\\

\begin{center} {\bf \small Appendix A: Magnetic spin moment} \end{center} 

\begin{figure}[b!]
 \includegraphics[scale=0.45]{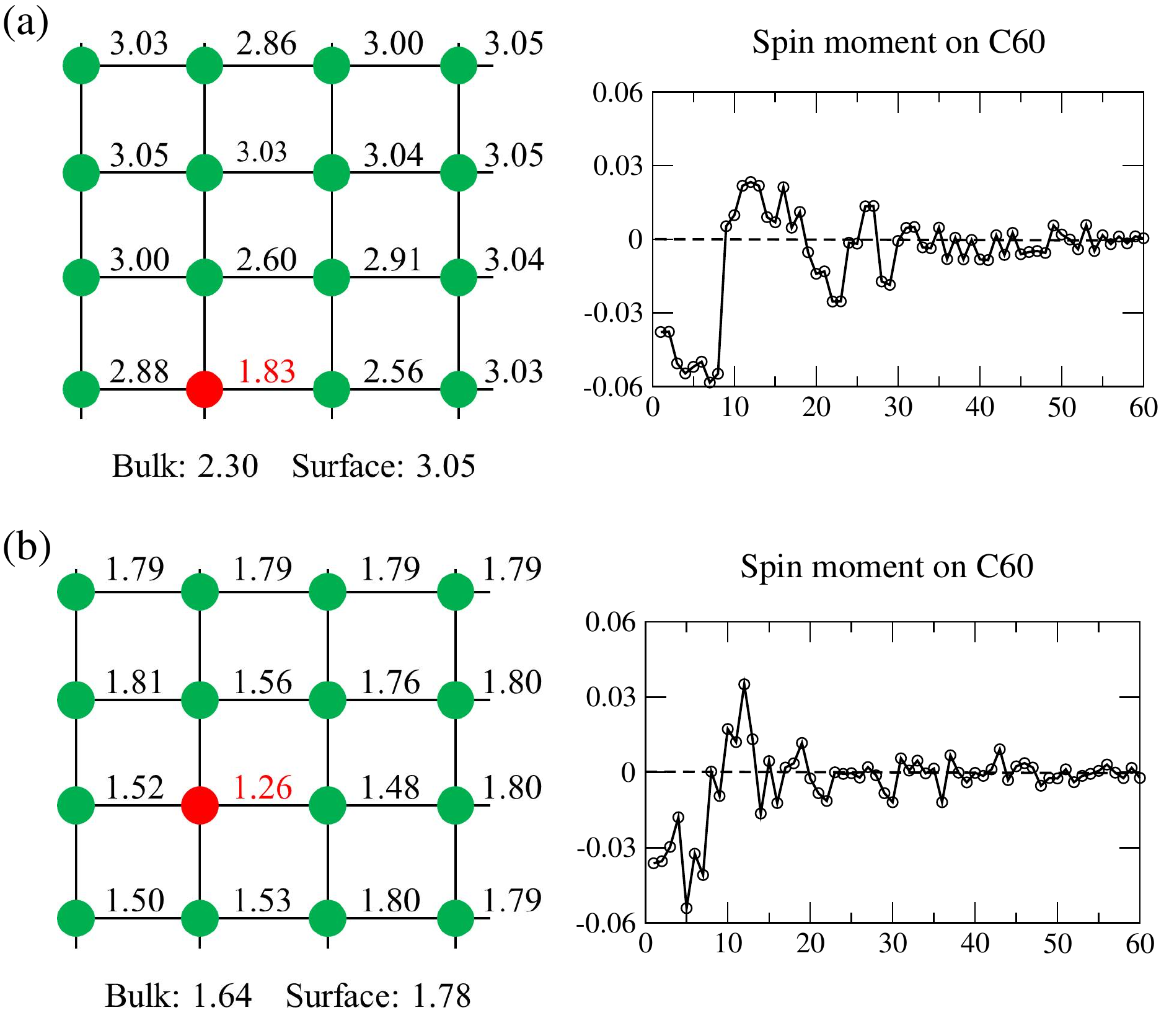}
 \caption{(Color online) Magnetic spin moment of (a) C$_{60}$/bcc-Fe(001) 6:6 bond, (b) C$_{60}$/hcp-Co(0001) 5:6 bond. Left panels: spin moment in $\mu_{\text{B}}$ of the surface atoms, the surface atom just below the bond is marked in red. The magnetic moment of bulk and clean surface are also indicated. Right panels: The distribution of induced spin moment in the C$_{60}$ with respect to the number of atoms, note that trajectory for numbering the atoms beginning from the carbon atoms at the interface and moving away from the interface. \label{spin_moment}}
\end{figure}

The calculated magnetic spin moment of C$_{60}$/bcc-Fe(001) and C$_{60}$/hcp-Co(0001) is shown in Fig. \ref{spin_moment} for the lowest energy configurations, a strong modification for both surface and adsorbate have been found with the concomitant interaction bewteen molecule and substrate. 

We present the spin moment of the first layer of surface (see the left panel of Fig. \ref{spin_moment}), in both Fe and Co surfaces we notice the signaficant reducement of the spin moment in the vicinity of molecule. This drop of spin moment in surfaces originates from the hybridization between $d$ states (particulally pronounced for the out-of-plane extended $d$ orbitals) of surfaces and $\pi$-molecular orbitals. In particular, the largest decrease of spin moment occurs for surface atom just below molecule (marked in red) of $M_{\text{s}}$ = 1.83 and 1.26 $\mu_{\text{B}}$ for C$_{60}$/Fe(001) and C$_{60}$/Co(0001), respectively, which means a drop of about 40 \% and 30 \% compared to the clean Fe(001) and Co(0001) surfaces. Additionally, the spin moment of this particular atom becomes even smaller than the bulk value of 2.30 and 1.64 $\mu_{\text{B}}$ for bcc-Fe and hcp-Co, respectively.

In addition, the induced magnetic moment of C$_{60}$ is polarized negatively and mainly localized around to the interface (see the right panel of Fig. \ref{spin_moment}). The spin moment appears to oscillate when moving away from the interface and converge to the expected value zero finally. We estimated the total magnetic moment of C$_{60}$ by summing over the local spin moments calculated by projecting the spin-resolved Kohn-Sham states onto the atomic wavefunctions, it was found to be of about -0.38 and -0.23 $\mu_{\text{B}}$ for C$_{60}$/Fe(001) and C$_{60}$/Co(0001), respectively. This transferred spin moment in the C$_{60}$ originates from the three electronic levels of LUMO close to the Fermi level.

\begin{center} {\bf \small Appendix B: Spin-polarized vacuum LDOS} \end{center} 

\begin{figure*}
 \includegraphics[scale=0.56]{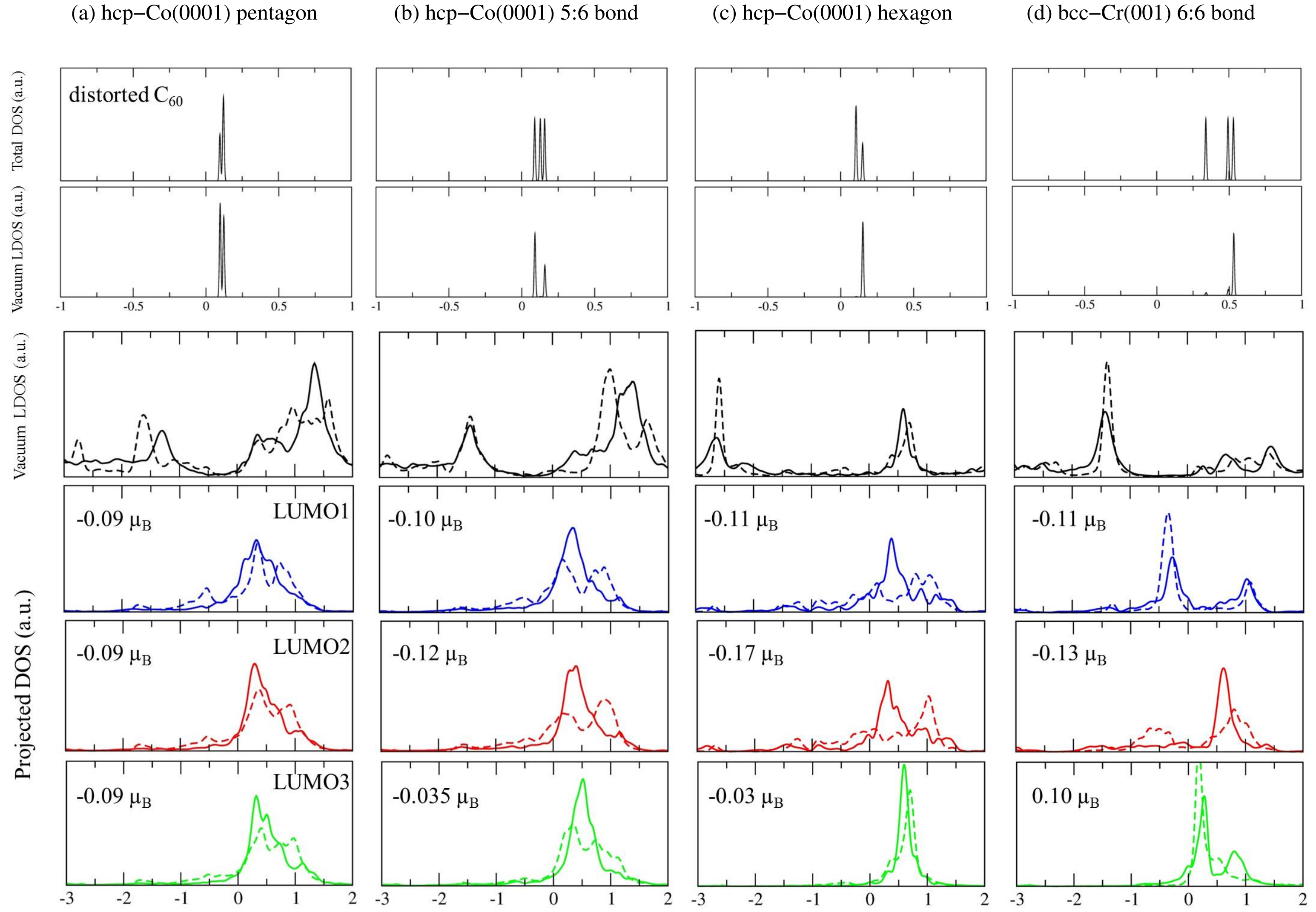}
 \caption{(Color online) Electronic structure of a C$_{60}$ molecule absorbed on top of hcp-Co(0001) with (a) a pentagonal ring, (b) 5:6 bond, (c) a hexagon ring geometry and (d) on top of bcc-Cr(001) with 6:6 bond. 
On upper panels the total DOS and the spin-resolved vaccum LDOS of the free (but distorted) C$_{60}$ molecules are shown. On lower panels the spin-resolved vaccum LDOS and PDOS on three LUMOs of the full C$_{60}$/ferromagnet system are presented. 
Spin up (down) data are plotted by solid (dashed)  lines. 
Spin polarization on each LUMO level is also indicated. \label{bilan_mag_surfaces}}
 \end{figure*}

We present in Fig. \ref{bilan_mag_surfaces} the spin-polarized DOS of a C$_{60}$ molecule absorbed on hcp-Co(0001) with three different absorption sites and on bcc-Cr(001) with 6:6 bond. Although the three LUMO orbitals are generally polarized we do not find any significant spin-polarization in the vacuum LDOS around the Fermi energy in all the cases. 

\begin{center} {\bf \small Appendix C: Adsorption geometry of C$_{60}$ on Co/Pt(111)} \end{center} 

\begin{figure}[b!]
 \includegraphics[scale=0.45]{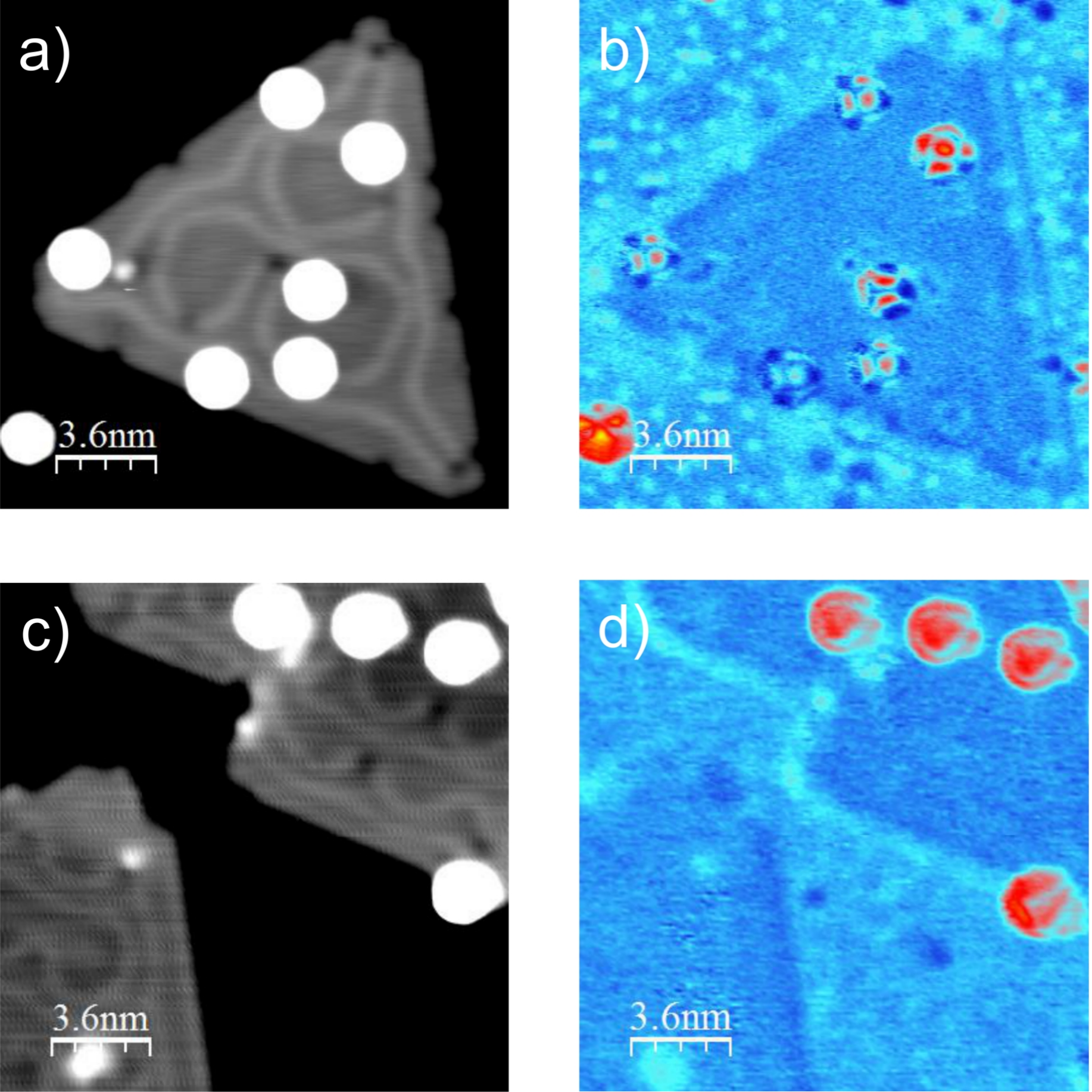}
 \caption{(Color online) $(18 \times 18)$~nm$^2$ STM images of single C$_60$ molecules (white round shapes) on Co/Pt(111) (a) topography image, as deposited at 4.6~K. (b) associated conductance image at the tunneling voltage 1~V corresponding to the experimental LUMO level. (c) topography image recorded at 4.6~K after an annealing of the sample at room temperature. (d) associated conductance image at the tunneling voltage 1~V, close to the LUMO level. (c) and (d) have been recorded with a Fe/W spin polarized tip, on the same molecules that are measured in Fig.~\ref{STS} (a).
 \label{C60_Co}}
\end{figure}

We present experimental details on the geometry of adsorption of single C$_{60}$ molecules on Co/Pt(111), as deposited on the sample at a temperature of 4.6~K [Fig.~\ref{C60_Co} (a) and (b)] and after an annealing of the sample at room temperature during few minutes [Fig.~\ref{C60_Co} (c) and (d)].  As can be seen in Fig.~\ref{C60_Co} (a) and (c), Co islands are quasi triangular and single layer high. One can observe brighter lines in between the C$_{60}$ molecules that have been ascribed to discommensuration lines\cite{meier_spin-dependent_2006}, separating different stacking area (Co in hcp or fcc stacking sites with respect to the Pt(111) surface layer). This makes the experimental system rather complex and prevents to ascribe a precise adsorption site for the C$_{60}$ molecules. Nevertheless, conductance images recorded at the energies of the different molecular states reveal unambiguously their adsorption geometry. Fig.~\ref{C60_Co} (b) is a conductance image recorded at 1~V that shows the local symetry of the different molecules. For example, three of them are nearly identical  in a vertex configuration. After an annealing at room temperature, Fig.~\ref{C60_Co} (c) shows that the molecules are surprisingly still isolated. A close look at the associated conductance image showing the LUMO state [Fig.~ Fig.~\ref{C60_Co} (d)] shows that the adsorption geometry has changed to pentagonal. Indeed, the dot and ring like structure observed in Fig.~\ref{C60_Co} (d) is very close the LUMO level observed and calculated for a pentagonal adsorption geometry on Cr(001)\cite{Alex-2012}. It is worth noting that the favored pentagonal adsorption geometry is very unusual on a close-packed surface and is probably due to the complex Co structure on Pt(111).

\bibliographystyle{apsrev}
\bibliography{biblio_C60_Fe_Co}

\end{document}